\documentclass{aa}
\usepackage{graphicx}
\usepackage{lscape}

\usepackage{times}
\usepackage{epsfig}
\usepackage{epsf}
\usepackage{rotating}

\begin{document}


\title{{\it SPEAR} far UV spectral imaging of highly ionized emission from the North Galactic Pole Region}


\author{
Barry Y. Welsh,\inst{1}
Jerry Edelstein,\inst{1}
Eric J. Korpela,\inst{1}
Julia Kregenow,\inst{1}
Martin Sirk,\inst {1}
Kyoung-Wook Min,\inst{2}
Jae Woo Park,\inst{2}
Kwangsun Ryu,\inst{2}
Ho Jin,\inst{3}
In-Soo Yuk \inst{3}
and Jang-Hyun Park \inst3}

\institute{Space Sciences Laboratory, University of California, 7 Gauss Way, Berkeley, CA 94720 \and
Korea Advanced Institute of Science $\&$ Technology, Dajeon, 305-70, Korea \and 
Korea Astronomy $\&$ Space Science Instutute, 305-348, Daejon, Korea}

\date{Submitted: April 2007}

\titlerunning{FUV emission}
\authorrunning{Welsh et al.}


\abstract
{}
{We present far ultraviolet (FUV: 912 - 1750\AA) spectral imaging
observations recorded with the $\it SPEAR$ satellite
of the interstellar OVI (1032\AA), CIV (1550\AA), SiIV (1394\AA), SiII* (1533\AA) and AlII (1671\AA)
emission lines originating
in a 60$^{\circ}$ x 30$^{\circ}$ rectangular region lying close to the North Galactic Pole.
These data represent the first large area, moderate spatial resolution maps of the
distribution of UV spectral-line emission originating the
both the highly ionized medium (HIM) and the warm ionized medium (WIM)
recorded at high galactic latitudes.}
{By assessing and removing a local continuum level that underlies these
emission line spectra, we have obtained interstellar emission intensity maps for the aforementioned
lines constructed
in 8$^{\circ}$ x 8$^{\circ}$ spatial bins on the sky.}
{Our maps of OVI, CIV, SiIV and SiII* line emission show the highest intensity levels
being spatially coincident with similarly high levels of soft X-ray emission originating
in the edge of the Northern Polar Spur feature. However, the distribution of the low
ionization AlII emission does not show this spatial correlation, and suggests that
warm-neutral and/or partially ionized gas with a temperature $<$ 20,000K may be quite pervasive at
high galactic latitudes.
 
The variation of the emission line intensity ratios
as a function of
sky position is contrasted with theoretical predictions
concerning the physical state of  interstellar gas
in the galactic halo. The observed line ratios alone unfortunately do not provide
us with a clear diagnostic tool to distinguish between the various physical production
mechanisms responsible for both high and low ion states. However, our results do
favor the hybrid model of Shull and Slavin (1994) which incorporates
the combined effects of turbulent mixing layers and isobarically cooling supernova
remnant gas.
For this highly ionized gas, our results 
are best explained assuming
that the observed OVI halo emission is somewhat clumpy in nature, consistent with its
 production at interfaces between warm (T = 10$^{3}$-10$^{4}$K) and hotter
 (T = 10$^{6}$K) soft X-ray emitting gas. CIV emission at these
interfaces occurs in the intermediate temperature  (T= 10$^{5}$K) gas, which
seems always present whenever OVI is strongly detected. Alternately, the data
are also consistent with
CIV emission being ubiquitous throughout the halo with
a fairly constant level of emission line intensity (of $\sim$ 4000LU), and our 
observations mostly reflect the superposition of spatially
separate OVI emission originating
at the cloud interfaces of random clumps of high latitude gas.}
{}


\keywords{ISM: emission: other  ---  ultraviolet: }

\maketitle

\section{Introduction}
The emission properties of the interstellar (IS) gas and
its associated spatial distribution on the sky are well known
at most wavelengths except for
the far ultraviolet region (912\AA\ - 2000\AA).
For example, emission from the cold (T $<$ 100K) interstellar
medium (ISM) is best
observed at radio and
millimeter wavelengths, particularly through 21cm mapping of atomic
hydrogen (Burton and Hartmann \cite{burton94}). In contrast, the
emitting properties of hot IS gas with temperatures $>$ $10^{6}$K
are best revealed through X-ray observations,
as shown by the all-sky emission maps recorded by
the $\it ROSAT$ satellite (Snowden et al. \cite{snow97}).
In the visible region, all-sky maps of H$\alpha$ emission have provided
us with a wealth of new insights into the spatial and kinematic structure
of the warm (T $\sim$ $10^{4}$K) and ionized component of  gas in our Galaxy 
(Haffner et al. \cite{haffner03}). However, although diffuse ultra-violet (UV) emission
from some very small regions
of the sky has been previously reported (Hurwitz \cite{hurwitz94}, Shelton et al. \cite{shelton01},
Murthy $\&$ Sahnow \cite{murthy04}), no large-scale sky-survey
in the UV had been
attempted
until the launch of the $\it SPEAR$ instrument in 2003
(Edelstein et al. \cite{edel06a}).
$\it SPEAR$, flown aboard the Korean STSAT-1, has now
provided us with the first UV spectral imaging
survey (912 - 1150\AA\ and 1350 - 1750\AA) of $\sim$ 80$\%$ of the sky
recorded at a spectral
resolution of $\lambda$/$\Delta$$\lambda$ $\sim$ 550 and an imaging resolution
of $\sim$ 5 arc min (Edelstein et al. \cite{edel06b}) 

Far UV photons ionize
interstellar (HI) gas atoms, dissociate interstellar
H$_{2}$ molecules and provide a significant contribution to the heating of the
ISM through the liberation of electrons from interstellar dust grains and by directly
exciting IS atoms and molecules (Black $\&$ van Dishoeck \cite{black87}). 
The far UV spectrum of diffuse interstellar emission contains the astrophysically important
cooling lines of CIV (1550\AA), SiIV (1394\AA) and OVI (1032\AA) for a typical IS plasma
with a temperature in the range T = $10^{4.5}$ - $10^{5.7}$ K (Korpela et al. \cite{korp06}).
In addition,
emission from the SiII* (1533\AA) and AlII (1671\AA) lines, which probe the
warm (T $\sim$ 10$^{4}$K)
and partially-ionized ISM (Jenkins $\&$ Tripp \cite{jenk01}),
are also prominent. Thus, spectral imaging
of this diffuse far UV emission can reveal the physical conditions, history, spatial distribution, cooling
physics and hydrodynamic processes of the galactic ISM.
The major component to the measured far UV diffuse flux in most galactic directions
is that of starlight (from hot OB stars), which
is scattered by interstellar dust (Bowyer \cite{bowyer91}). 
Thus, $\it SPEAR$ spectral maps of regions in and near the galactic plane correlate well
with those of  reddening and HI column density and are thus complimentary
to those recorded by the $\it IRAS$ and $\it COBE$ missions (Finkbeiner, Davis $\&$ Schlegel
\cite{fink99}).
However, at high latitudes in the galactic halo the diffuse far UV
emission signal, although far weaker than
that recorded at lower galactic latitudes,
is essentially uncontaminated by the combined effects of scattered UV photon
emission from dust, hot stars and interstellar
molecules. 

The detailed spatial distribution of the highly ionized gas in the halo, as
traced by the important UV lines of CIV, SiIV and OVI, is not known in great detail since such
data has been gained mainly from absorption measurements along individual
sight-lines to a limited number of distant OB stars and AGN. However, the recent
$\it FUSE$ absorption survey of
OVI absorption in the halo has revealed large irregularities in the distribution of highly
ionized gas that suggests a significant amount of both small- and large-scale structure in
this ionized medium (Savage et al. \cite{savage03}). Furthermore, in a recent
survey of diffuse OVI emission along 183 sight-lines with the $\it FUSE$ satellite,
Dixon, Sankrit $\&$ Otte (\cite{dixon06}) have found that the OVI emitting regions at high galactic
latitudes are intrinsically fainter than those at low latitudes, which may indicate two 
different populations of highly ionized emitting gas. Although most of the OVI emission line-intensities
lie in the  fairly restricted range of 1800 to 5500 LU (line unit; 1 photon cm$^{-2}$ s$^{-1}$
sr$^{-1}$ at 1032\AA), the nearby local emitting OVI regions have higher electron densities
and far smaller path lengths than regions emitting at high galactic latitudes.
 
The ions of CIV and SiIV commonly trace IS gas with a
lower temperature of $\sim$ 10$^{4.5-5.0}$K. In a study of high
ion absorption
through the galactic halo by Savage, Sembach $\&$ Lu (\cite{savage97}), the
authors found that
the column density ratio of N(CIV)/N(SiIV) was relatively constant (with
a value of 4.2) for most sight-lines irrespective of the galactic latitude being probed.
However, there is growing evidence that the column density ratio of N(CIV)/N(OVI) in the
galactic halo does not follow such behavior and its measured value along
a particular halo sight-line may well reflect the local nature of ionization
processes present (Zsargo et al. \cite{zsargo03}), Savage et al. \cite{savage03}). This
may be expected if UV radiation from the
central stars of OB associations and stellar wind-driven bubbles is escaping into
the halo and producing the CIV and SiIV ions, but is not contributing
to the emission from the higher ionization
species of OVI and NV ions which are
formed
mainly by collisional ionization processes (Ito $\&$ Ikeuchi \cite{ito88}). 
Unfortunately studies of the important CIV and SiIV lines observed in emission
have proven problematic
for observers (see Bowyer \cite{bowyer91} for a review). Apart from the detections
of diffuse emission with line intensities
of $\sim$ 2000 - 5700 LU from the lines of CIV and OIII along
several sight-lines by Martin and Bowyer (\cite{martin90}), no
other similar emission measurements had been made during
the intervening years until the launch of the
$\it SPEAR$ instrument in 2003 (Edelstein et al. \cite{edel06a}).

Thus, in order to explore the spatial distribution of 
10$^{4.0}$ - 10$^{6}$K emitting gas
in the halo we present
$\it SPEAR$ spectral imaging data of the interstellar OVI, CIV, SiIV,
SiII* and AlII emission lines originating
in rectangular region of size  $\sim$ 60$^{\circ}$ x 30$^{\circ}$
lying close to the North
Galactic Pole. This high latitude region
is largely devoid of bright and hot OB stars (that can contaminate
the observed far UV emission signal), 
and has a relatively low value of interstellar reddening ($<$ 0.05) and
is located far away from the effects of
scattered starlight from the underlying galactic plane regions. Our $\it SPEAR$
observations of the
the OVI, CIV and SiIV emission lines thus provide us with a potential probe of the
the hot (10$^{5 - 6}$K) and highly ionized medium (HIM), whereas the SiII* and AlII lines
trace the cooler (10$^{4}$K) gas of the warm and partially-ionized medium (WIM). 
These FUV data are presented
in the form of spectral line image maps binned 
into 8$^{\circ}$ x 8$^{\circ}$ regions, with an initial instrumental spatial resolution
of $\sim$ 15 arc min.
 
\section{Observations and Data Reduction}
We present observations of diffuse emission from near the North Galactic Pole region recorded with
both the
long wavelength (L-band: 1350 - 1750\AA)
and short wavelength (S-band: 912 - 1150\AA) channels
of the $\it SPEAR$ imaging UV spectrograph
flown aboard the Korean STSAT-1 satellite, which was launched in September 2003 (Edelstein
et al. \cite{edel06a}, \cite{edel06b}). An area of sky covering
a rectangular strip of size $\sim$ 60$^{\circ}$ x 30$^{\circ}$ centered on ($\it l$ = 280$^{\circ}$,
$\it b$ = +70$^{\circ}$) was observed with repeated scans using the
4$^{\circ}$ x 4.3$\arcmin$ S-band and 7.5$^{\circ}$ x 4.3$\arcmin$ L-band
entrance apertures, which resulted in a total on-sky exposure time
of 32ksec obtained from $\sim$ 300 orbital scans. 

Full details of the $\it SPEAR$ scientific mission,
its instrument design and modes of on-orbit observation 
can be found in Edelstein et al. (\cite{edel06a}, \cite{edel06b}). 
The
standard data reduction procedures for $\it SPEAR$ observations are
described in detail by Korpela et al. (\cite{korp06})
and consist of 4 main elements that include: (i) re-mapping of each
time-tagged FUV photon to a position
on the sky based on knowledge of the spacecraft's time
and attitude information, (ii) correction for the effective
exposure level for each sky-pixel, (iii) removal of stellar contamination from
the collected photons by excluding
spatial bins with $>$
3 times the local median count rate, and (iv) binning the resultant data to an effective spatial resolution
of $\sim$15$\arcmin$ pixels on the sky.  Each of these sky-pixels can thus be
associated with a data cube of information containing
the  FUV emission spectrum (i.e. wavelength versus counts s$^{-1}$) summed over the area of a 
sky-pixel at a particular galactic position ($\it l$, $\it b$) on the sky. Thus,
the resultant data product
from such observations is a far UV spectral image for each sky-pixel. 

In Figures 1 and 2 we show the total summed UV spectra
(for the respective S- and L-bands) from all of the sky-pixels contained
within the $\sim$ 60$^{\circ}$ x 30$^{\circ}$ area of the sky presently under investigation. These summed spectra, which
contain contributions from areas of both intrinsically high and low FUV signal, have
been binned to a spectral resolution of 1.5\AA\ (S-band) and 3.0\AA\ (L-band). As
can be clearly seen in Figure 2,
the dominant component
to the measured flux at longer FUV
wavelengths is from the underlying continuum, which consists of contributions
from the detector background, dust-scattered continua and scattered airglow within the
instrument.  Although not a true continuum signal, this underlying  level corresponds
to $\sim$ 950 CU (continuum intensity units; photons s$^{-1}$ cm$^{-2}$ sr$^{-1}$ \AA$^{-1}$) at 
a wavelength of 1550\AA.
At shorter wavelengths the underlying continuum contribution
from scattering by dust is somewhat reduced and the dominant source of emission arises
in scattering from the Lyman series of geo-coronal hydrogen lines. As a guide, this underlying
signal is $\sim$ 3300 CU at 1030\AA. 
However, our main interest in this Paper is the overlying (far fainter) astrophysical emission
lines of OVI, CIV, 
SiIV, SiII* and AlII formed in the diffuse
ISM. We also note the low level of
emission
from molecular H$_{2}$ fluorescence lines (at 1608\AA), which Edelstein et al. (\cite{edel06a}) have detected
ubiquitously at galactic latitudes $<$ 50$^{\circ}$. This can be
explained by the presence of
very low levels of interstellar dust found at the presently observed high galactic latitudes.

\begin{figure*}
\center
{\includegraphics[width=12cm]{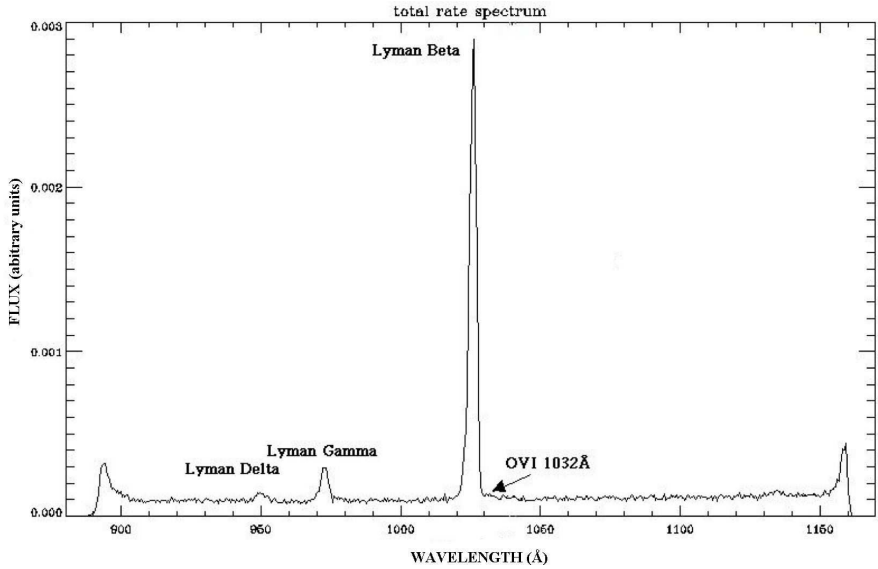}}
\caption{Summed far UV S-band (900 - 1150\AA) $\it SPEAR$ spectrum of the $\sim$ 60$^{\circ}$ x 30$^{\circ}$ area of sky near to the North Galactic Pole region. The underlying `continuum' level amounts
to $\sim$ 3300 CU at 1030\AA. See Table 1 for total summed line intensity values.}
\label{Figure 1}
\end{figure*}

\begin{figure*}
\center
{\includegraphics[width=12cm]{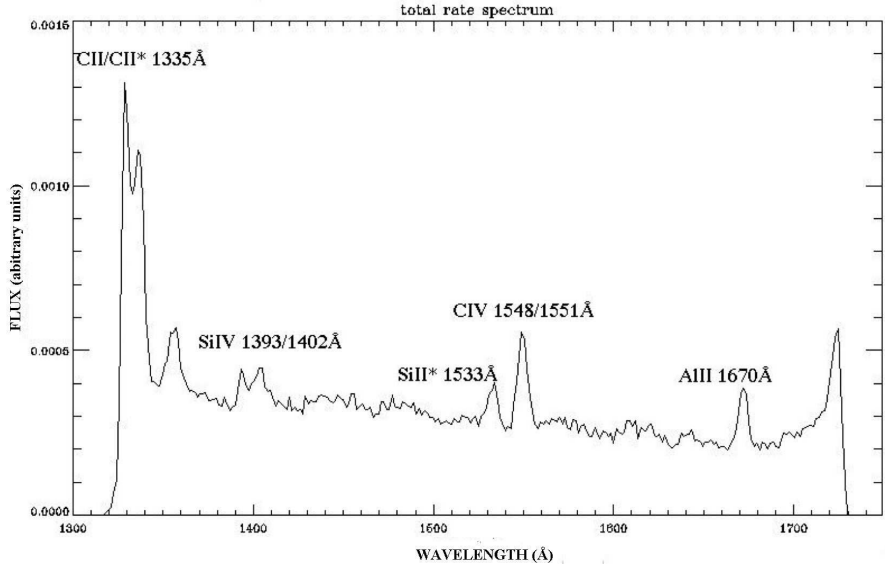}}
\caption{Summed far UV L-band (1350 - 1750\AA) $\it SPEAR$ spectrum of the $\sim$ 60$^{\circ}$ x 30$^{\circ}$ area of the sky near to the North Galactic Pole region. The underlying `continuum' level
is $\sim$ 1000 CU at 1550\AA. See Table 1 for total summed line intensity values.}
\label{Figure 2}
\end{figure*}

In order
to improve the resultant S/N ratio of the data, the 15 arc min sky-pixels were
binned into far larger 8$^{\circ}$ x 8$^{\circ}$ elements to obtain a summed,
exposure corrected
FUV spectrum over areas of 64 sq. deg. on the sky.
For each of the
emission lines of interest in these large sky-pixel FUV spectra, an underlying
continuum level (modeled using a dust scattered
population of upper main sequence stars as described in Korpela et al. \cite{korp06}) was assessed over a wavelength
range extending $\sim$$\pm$10\AA\  from the expected central position of the
spectral line. As part of
this continuum assessment a best-fit gaussian emission profile (superposed
upon this level) was also attempted for any emission
feature within $\pm$1.5\AA\ (i.e. half of one $\it SPEAR$ resolution element) from the theoretical
line center wavelength. This is an iterative process that seeks
to simultaneously best-fit  both the underlying continuum level and
the overlying gaussian profile, such that the result is a residual intensity (in counts sec$^{-1}$ or LU)
of the central emission feature together with its model fit parameters and their associated
$\chi$$^{2}$ fitting errors and a measure of the the detection significance
with respect to the level of noise in the data (both with and without the presence of
the emission
feature of interest). 
The fit procedure also takes into account the limited
resolution of the $\it SPEAR$ data, such that all interstellar emission lines
with wavelengths close to the line of interest are also fit simultaneously.
In the case of the CIV emission lines at 1548\AA\ and 1551\AA\ the observed profile
was fit simultaneously as a doublet feature and also taking into account the
presence of the nearby SiII* emission line.
Because of the close proximity of the strong Lyman beta airglow feature,the
O VI lines at 1032\AA\  and 1038\AA\ were fit
with empirical emission line profiles (constructed from the Lyman beta line) instead of a
pure Gaussian profile.
The underlying continuum background for
these lines was determined empirically by adding many spectra where
O VI was not detected. The resulting model was thus the sum of Lyman Beta,  the two O VI
features, the C II features (1036.34 and 1037.02 \AA), and the continuum background.
The free parameters of the fit were the intensities of the various emission lines, the intensity of
the background, and the wavelength centroid of the Lyman Beta line.
The wavelengths of the other features were kept as constant offsets relative to the Lyman
Beta line. 

\begin{figure*}
\center
\includegraphics[width=15cm,height=8cm]{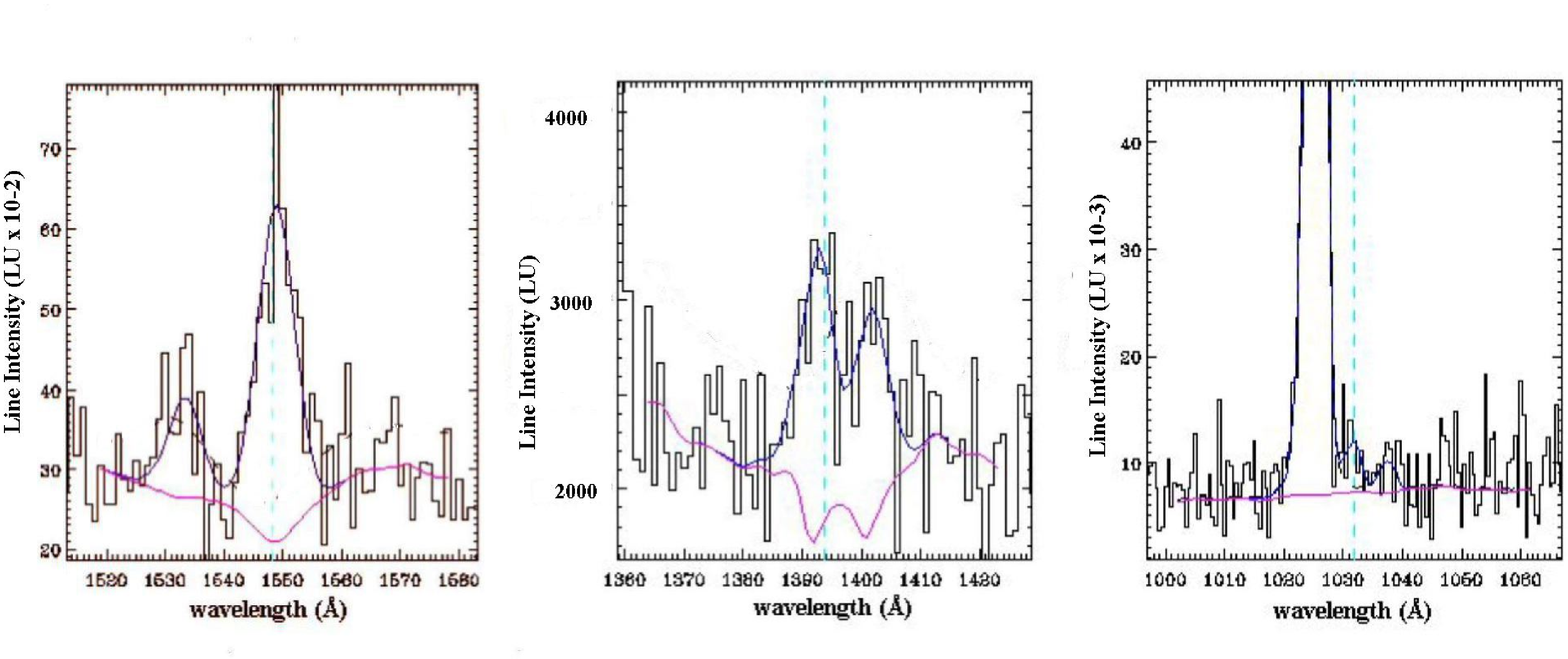}
\caption{Typical emission line fits to the CIV doublet at 1550\AA, the SiIV doublet
at  1394\AA\ and
the OVI 1032\AA\ doublet. The lower full line represents the best-estimate of the underlying
dust-scattered stellar continuum.}
\label{Figure 3}
\end{figure*}

In Figures 3 (a) - (c)
we show typical examples to the quality of the fits to some of the line spectra for
several of the (large) sky-pixels.
The final result of this line-fitting process are 2-D maps (in galactic co-ordinates)
of the emission line intensities (in LU) contained within spatial bins of
64 sq. deg. on the sky. In Figures 4 and 5 we show  the
relevant line-maps of the
CIV (1550\AA\ doublet), SiIV (1394\AA), OVI (1032\AA\ doublet), SiII* (1533\AA) and
AlII (1671\AA) 
emission (in LU) as a function of position on the sky.
In these plots we have labeled 16 of the 8$^{\circ}$ x 8$^{\circ}$ sky-pixels
with letters from A to P, with their
associated area averaged emission line intensities 
being listed in Table 1 together with their associated measurement errors. We note
the quite different contributions to the overlying
emission line intensities from the underlying model stellar continua for each line.
In general we report line detections and fits with a confidence level of
$>$ 3$\sigma$ significance, but in order to present results for the weakly detected OVI line we have reduced the confidence level for this line to detections with $>$ 2$\sigma$ significance.
We also note that the
interstellar SiIV line at 1394\AA\ is blended
with that of both SiIV (1402\AA) and OIV] (1400\AA) emission, which Korpela et al. (\cite{korp06}) have estimated
to possess a comparable emission intensity. Therefore, since we do not know the exact
contribution from OIV to the SiIV (1402\AA) line, we shall assume a 2:1 ratio for the SiIV
doublet components (based on their oscillator strengths), and thus we report 1.5 times the measured
value of the SiIV (1394\AA) line as being representative of the total doublet value for that ion species.
Similarly we have scaled the measured intensity of the (stronger) OVI (1032\AA) line to
report a value for the OVI doublet, based on the oscillator strength of the
weaker OVI (1038\AA) line. The intensity value for CIV reported in Table 1 is for the fitted
line-doublet, whereas the emission levels reported for both SiII* and AlII are for the single
lines only. Finally, in the last row of Table 1 we show (for comparison purposes) the equivalent emission line data for
the North Ecliptic Pole region as measured with $\it SPEAR$ (Korpela
et al. \cite{korp06}). 

It should be noted that we have made no attempt to correct the line intensity values listed in
Table 1 for extinction by interstellar dust. Figure 4(d) shows that the
general level of extinction is low in most sky-pixels (i.e. E(B-V) $<$ 0.05 mag).
However, it should be noted that even this (small) amount of dust can reduce the flux
at the OVI 1032\AA\ line by a factor of 2.
Since the precise
location of the source of UV line emission with respect to
dust absorption is not well-known from our 8$^{\circ}$ x 8$^{\circ}$ maps, we can only refer the reader
to the extinction values shown in Figure 4(d) for an estimation of the possible levels of
emission line attenuation for each sky-pixel.

\begin{landscape}
\begin{table*}
\begin{center}
\caption{$\it SPEAR$ measured FUV emission line intensities (in LU) averaged over 8$^{\circ}$ x 8$^{\circ}$ bins at sky positions A - P.}
\begin{tabular}{lccccccccc}
\hline
\hline
$\bf Sky Pixel$&$\bf l$&$\bf  b$&$\bf I(OVI)$ &$\bf I(CIV)$&$\bf I(SiIV)$&$\bf I(SiII*)$&$\bf I(AlII)$&$\bf I(OVI)/(CIV)$&$\bf I(CIV)/I(SiIV)$\\
&&&(scaled doublet)&(doublet)&(scaled doublet)&&&&\\
&&&(LU)&(LU)&(LU)&(LU)&(LU)&&\\
\hline
A&301.5$^{\circ}$&+62.8$^{\circ}$&21,160$\pm$7500&7,840$\pm$660&3,530$\pm$610&2,995$\pm$655&6,035$\pm$1780&2.7$\pm$0.98&2.2$\pm$0.55\\
B&312.8$^{\circ}$&+71.5$^{\circ}$&$<$ 6,300&7,770$\pm$580&4,290$\pm$2185&1,830$\pm$690&4,230$\pm$1080&$<$0.8&1.8$\pm$1.0\\
C&341.6$^{\circ}$&+78.8$^{\circ}$&16,420$\pm$8000&7,850$\pm 310$&2,610$\pm$300&2,580$\pm 335$&6,335$\pm$2090&2.1$\pm$1.1&3.0$\pm$0.48\\
D&303.8$^{\circ}$&+78.5$^{\circ}$&$<$6,200&6,890$\pm$405&2,215$\pm$420&2,355$\pm$455&3,400$\pm$1130&$<$0.9&3.1$\pm$0.71\\
E&292.0$^{\circ}$&+69.1$^{\circ}$&10,320$\pm$5500&6,350$\pm$450&2,490$\pm$765&1,925$\pm$445&3,090$\pm$1515&1.6$\pm$0.96&2.6$\pm$0.98\\
F&287.5$^{\circ}$&+59.7$^{\circ}$&$<$6,300&8,105$\pm$530&3,330$\pm$485&2,200$\pm$430&6,285$\pm$1200&$<$0.8&2.4$\pm$0.50\\
G&276.3$^{\circ}$&+55.4$^{\circ}$&15,830$\pm$6500&6,515$\pm$455&2,365$\pm$630&1,475$\pm$410&4,930$\pm$1290&2.4$\pm$1.1&2.8$\pm$0.92\\
H&276.3$^{\circ}$&+64.9$^{\circ}$&9,350$\pm$4200&4,695$\pm$385&2,140$\pm$385&1,860$\pm$405&4,370$\pm$920&2.0$\pm$1.0&2.2$\pm$0.57\\
I&276.3$^{\circ}$&+74.7$^{\circ}$&$<$4,000&3,670$\pm$365&1,565$\pm$370&1,320$\pm$320&3,860$\pm$1525&$<$1.1&2.3$\pm$0.76\\
J&276.2$^{\circ}$&+84.8$^{\circ}$&6,765$\pm$3900&5,775$\pm$225&1,830$\pm$220&1,910$\pm$315&4,260$\pm$1470&1.2$\pm$0.7&3.2$\pm$0.50\\
K&248.7$^{\circ}$&+78.5$^{\circ}$&10,000$\pm$5000&4,390$\pm$345&2,140$\pm$700&1,435$\pm$295&4,700$\pm$1310&2.3$\pm$1.3&2.1$\pm$0.84\\
L&260.6$^{\circ}$&+69.1$^{\circ}$&$<$5,050&3,175$\pm$520&1,975$\pm$510&2,615$\pm$470&3,595$\pm$920&$<$1.6&1.6$\pm$0.67\\
M&265.1$^{\circ}$&+59.7$^{\circ}$&$<$6,000&4,715$\pm$645&2,855$\pm$840&1,540$\pm$470&6,020$\pm$1525&$<$1.3&1.7$\pm$0.73\\
N&251.1$^{\circ}$&+62.8$^{\circ}$&$<$6,200&2,035$\pm$730&$<$1200&$<$600&$<$2,500&$<$3.0&$>$1.7 \\
O&239.8$^{\circ}$&+71.5$^{\circ}$&$<$6,580&4,455$\pm$570&$<$1400&2,300$\pm$550&3,275$\pm$840&$<$1.5&$>$3.1\\
P&211.1$^{\circ}$&+78.7$^{\circ}$&$<$7,500&4,150$\pm$600&$<$1300&1,050$\pm$505&3,315$\pm$825&$<$1.8&$>$3.2\\
N.E.P.**&96.0$^{\circ}$&+29$^{\circ}$&5725$\pm$570&5820$\pm$280&1430$\pm$120&2430$\pm$200&N/A&0.98$\pm$0.15&4.1$\pm$0.53\\
\hline
\hline
\multicolumn{10}{l}{ ** = observations of the North Ecliptic Pole (NEP) region by Korpela et al. (2006)} \\
\hline
\end{tabular}
\end{center}
\end{table*}
\end{landscape}

\section{Discussion}
\subsection{The CIV emission map}
Figure 4(a) clearly shows that the greatest line-emission intensity from this ion is found along
the left-hand edge of the region under investigation (i.e. the strip of sky
with galactic longitude $\it l$ $>$ 290$^{\circ}$).
The emission
intensity along this edge (within sky-pixels A to F) is
typically  $>$ 6500LU, which is to be compared with that of
$<$ 4500LU for the remainder of the region (sky-pixels K to P).
These emission levels are numerically consistent with those observed by
Martin and Bowyer (\cite{martin90}) who found CIV line (doublet) intensities in the range
2200 - 5700LU for similarly high latitude halo gas.

\begin{figure*}
\center
\includegraphics[width=15cm]{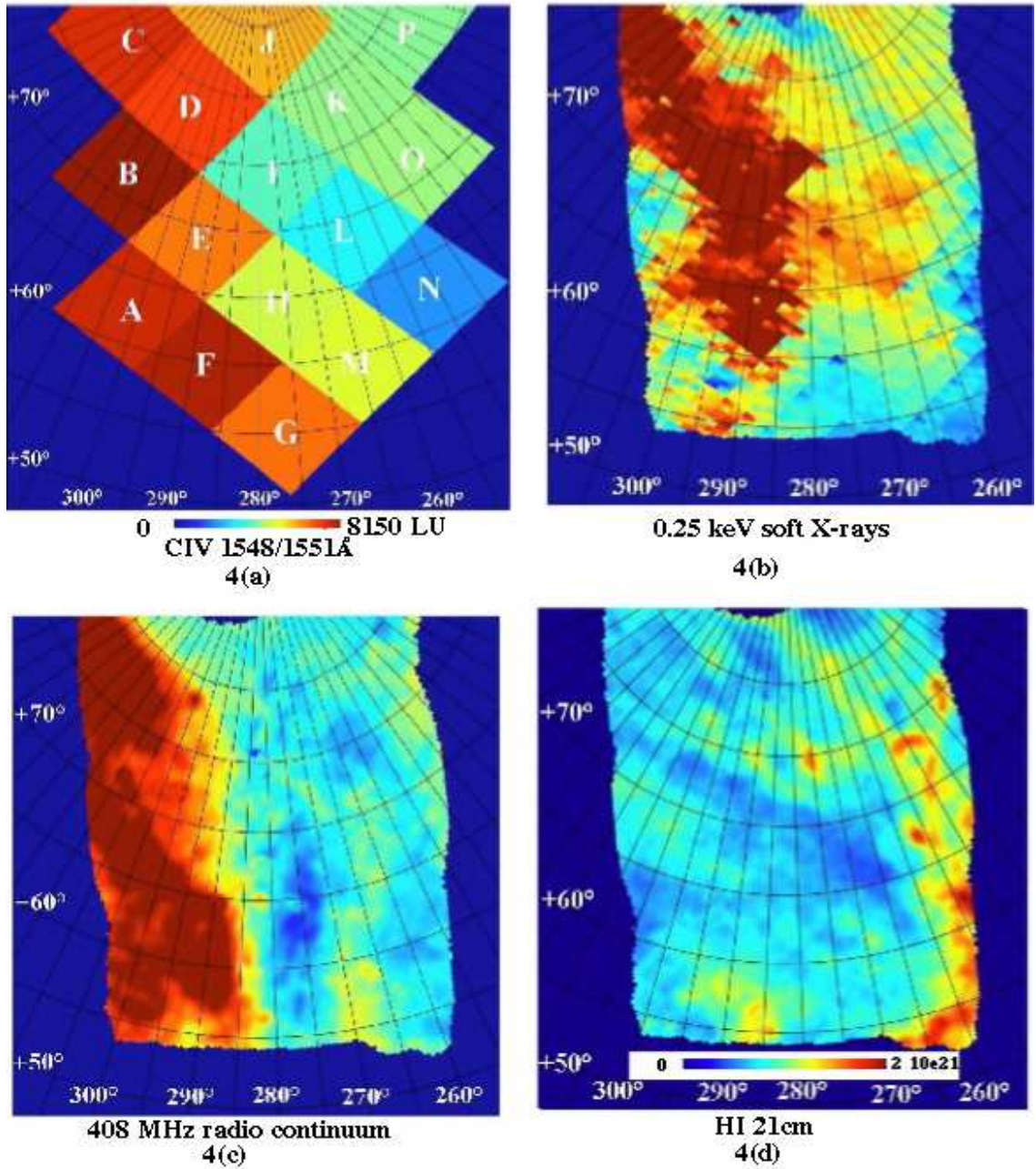}
\caption{4(a): SPEAR CIV emission line intensity map for the North Galactic Pole region. The highest levels of emission (shown in dark red) are found in sky-pixels labeled A, B, C, D and F, which are also spatially coincident with the high levels of 0,25 keV soft X-ray emission from the edge of North Polar Spur feature shown in 4(b). Maps of  408MHz radio continuum
and HI 21cm radio emisison from the same region of the sky are also shown in Figures 4(c) and 4(d).
All emission intensity scales are linear. }
\label{Figure 4}
\end{figure*}

In Figure
4(b) we show the same region
of the sky recorded at 0.25keV soft X-ray wavelengths 
with the $\it ROSAT$ all-sky survey (Snowden et al. \cite{snow97}), and similarly
in Figure 4(c) we show the 408MHz radio continuum emission (Haslam et al. \cite{haslam82})
and in Figure 4(d) the corresponding 21cm HI radio emission (Hartmann and Butler \cite{hart97}).
We note that this latter map is very similar to
that of the distribution of  dust IR emission as recorded by the $\it COBE$ and $\it IRAS$ satellites (Schlegel,
Finkbeiner $\&$ Davis \cite{schleg98}).
These three maps have been constructed using
the HEALpix routines of  Gorski et al. ( \cite{gorski05}), typically with spatial
resolutions $<$ 10 arc minutes.
A comparison of these maps with that of the CIV line-intensity
distribution in Figure 4(a)
clearly reveals three main aspects: (i) there is a high spatial correlation between regions
of high CIV line-emission intensity and high intensity values of
408 MHz radio continuum emission, (ii) the region of high CIV line-emission
intensity lies adjacent to, but not completely co-aligned with, that 
of the highest levels of soft X-ray emission, and (iii) there is an
anti-correlation between regions of high gas and dust emission and those of
high CIV, high 408 MHz and high soft X-ray emission intensity. 
This latter result is not unexpected, since even
small amounts of neutral gas and interstellar dust 
can easily shadow or totally absorb soft X-ray photon and
408 MHz emission from fast moving electrons, in addition to
efficiently scattering UV photons.

\begin{figure*}
\center
\includegraphics[width=15cm]{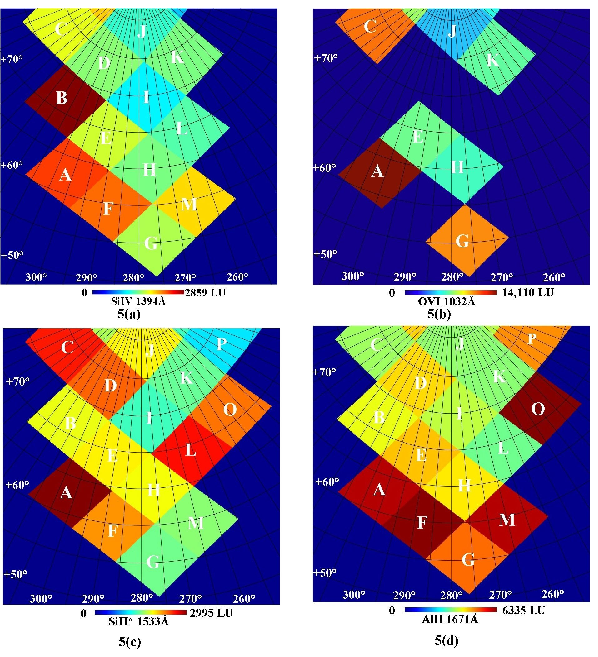}
\caption{SPEAR line emission maps for : (a) SiIV 1394/AA/ emission, (b) OVI 1032/1037/AA/,
(c) SiII* 1533/AA/ and (d) AlII 1671\AA. These maps are directly comparable with those shown in Figure 4 for the same region of the sky.}
\label{Figure 5 }
\end{figure*}

Willingdale et al. (\cite{willing03}) have analyzed the soft X-ray background signal
in this direction and found that the NPS and halo emission components lie behind
at least 50 per cent of the line-of-sight cold gas. The cold (HI) gas clouds are thought
to be located at distances over
the range 120 - 200pc in this general direction, as determined by both
photometry (Haikala et al. \cite{haik95}) and NaI absorption studies
towards nearby early-type stars (Lallement
et al. \cite{lall03}).  The largest concentration of HI (and dust) clouds
is located at galactic longitudes $<$ 265$^{\circ}$,
with an estimated neutral
interstellar HI column density of $>$ 10$^{20}$cm$^{-2}$, whereas the remainder
of the area is composed
of numerous filamentary clouds with a typical HI column density
$<$ 6 x 10$^{19}$ cm$^{-2}$ (based on NaI observations towards the star HD 109860
of distance 200pc, Lallement et al. \cite{lall03}). We also note that
the reddening towards most of this region has a low value of E(B-V) $<$ 0.05.
Since the region of low-level CIV emission ($\it l$ = 265 - 275$^{\circ}$, sky-pixels
K  to  P) 
is spatially coincident with that of significantly reduced soft X-ray emission, it is
reasonable to assume (like Willingdale et
al. \cite{willing03}) that 
a significant proportion of this  diffuse ionized gas emission lies behind the foreground
neutral HI gas clouds at distances $>$ 200pc in the overlying halo. Such gas can be
identified with that commonly traced by absorption studies of the
CIV and SiIV ions detected routinely towards halo
stars with distances of $\it z$ $<$ 4 kpc (Savage, Sembach $\&$ Lu \cite{savage97}).

The NPS feature has been conjectured to be a nearby supernova
remnant (Berkhuijsen, Haslam and Salter \cite{berk71}), or an ionized outflow from 
stellar winds of the underlying Sco-Cen OB association (Egger and Aschenbach \cite{egger95}).
We note that at high galactic latitudes the NPS, 
as seen both in radio and X-ray emission, appears as
a limb-brightened shell, as may be expected from the perspective of an expanding
supernova remnant. Our presently discovered enhancement of CIV line-emission
is seen in regions facing towards the (lower galactic latitude) NPS, for
galactic longitudes $>$ 285$^{\circ}$. This picture would seem to support
the notion that the region of the highest intensity
408 MHz radio continuum emission (which is caused by synchrotron processes)
is associated with shocks that delineate the outer shell
of the NPS and are also contributing to the observed high levels of CIV emission. The region
with galactic longitudes $<$ 285$^{\circ}$ lies beyond the giant NPS feature and is probably
representative of the general diffuse halo gas with a lower associated CIV emission line
intensity of $\sim$ 4000LU.

\subsection{The SiIV emission map}
Figure 5(a) shows the spatial distribution of SiIV (1394\AA) line emission, with the highest emission levels
occurring in sky-pixels A, B and F with intensities $>$ 3,300LU. Apart from sky-pixel M, the remainder
of the region is characterized by lower-level SiIV emission typically with a value $<$ 2,600 LU. We note that although sky-pixels A, B and F are also associated with similarly high levels of CIV line emission, 
the spatial distribution of SiIV emission is not as well-correlated with the highest levels of both
0.25 keV soft X-ray and 408MHz radio emission that trace the outer
edge of the NPS (shell) feature. We note
that SiIV emission
is normally associated with warm and ionized gas (often the
sites of small-scale mixing layers or conductive
thermal fronts) at a temperature of $\sim$ 60,000K, in
contrast with the far higher temperature (T $\sim$ 10$^{6}$K) and
more wide-spread soft X-ray emitting regions. The very lowest levels of SiIV emission
were recorded as upper limits ($<$ 1400LU) in the sky-pixels with galactic longitudes
$<$ 270$^{\circ}$. We confirm the notion (forwarded in the previous section)
that such regions are unconnected with the NPS shell feature
and presumably associated with general halo gas emission. This is in accord with the
observations of the North Ecliptic Pole (NEP) region with $\it SPEAR$ by Korpela el al. (\cite{korp06}),
who measured an emission level of 1430LU for the SiIV doublet. 

\subsection{The OVI emission map}
Figure 5(b) shows the spatial distribution of the OVI (1032\AA) line emission. Due to both the relative
weakness of emission from this line and the associated difficulty in its detection due to the proximity
to the strong
geocoronal Lyman Beta line, firm detections were made in only 7 of the 16 sky-pixels. The highest
observed emission levels were found in sky-pixels A and C, which can be associated with
positions of the highest CIV, soft X-ray and 408MHz emission within the edge of the NPS feature.
Our detection lower limit (for the OVI doublet) is $\sim$ 6200LU, which is marginally above the level
of 5725LU measured in extended $\it SPEAR$
observations of the NEP region by Korpela et al. (\cite{korp06}). We note that regions
from which we have been unable to measure detectable levels of OVI emission
(e.g. sky-pixels B, F,M,N and O) can apparently be associated with the presence
of foreground HI gas clouds shown in Figure 4(d). Such clouds may account
for the clumpy nature of the observed distribution of the OVI ion.

Emission from the OVI line is generally associated with highly ionized gas
in collisional equilibrium with a temperature
of $\sim$ 300,000K. Such gas cools rapidly and thus this ion traces interstellar regions
in transition either through cooling or at the interfaces between warm and
hotter gas. Such sites are prime regions for the formation of both CIV and SiIV ions at
a temperature of $\sim$ 10$^{5}$K. In a recent survey of diffuse OVI emission by
Dixon, Sankrit and Otte (\cite{dixon06}) using the $\it FUSE$ satellite, they observed
5010LU of OVI (1032\AA) emission from a region
of area 30 arcsec x 30 arcsec ($\it l$ = 284$^{\circ}$, $\it b$ = +74.5$^{\circ}$), which
is comparable to the average level of $\sim$ 6,800 LU (for the OVI 1032\AA\ line),
scaled from observations of the OVI doublet recorded in the 8$^{\circ}$ x 8$^{\circ}$
sky-pixels E $\&$ K by $\it SPEAR$. For comparison
purposes, the associated level of soft X-ray flux at this position
on the sky is 4520 x 10$^{-6}$ counts
s$^{-1}$ arcmin$^{-2}$ (Snowden et al. \cite{snow97}). 

\subsection{The SiII* emission map}
Figure 5(c) shows the spatial distribution of emission from the SiII* (1533\AA) line,
which primarily exists over an ionization potential range of 8.2 to 16.3eV and thus traces
ionized and partially ionized/neutral gas regions. Under collisional ionization
conditions its maximum emission occurs at a temperature of $\sim$ 12,500K.
The present 
non-detection of the nearby SiII (1526\AA) line (see
Figure 2) is attributed to opacity
effects (Korpela et al. \cite{korp06}, Shinn et al. \cite{shinn06}). Although the highest
emission levels
of SiII* ($\sim$ 2500LU) are
to be found in sky-pixels A to F (which are spatially coincident with significant soft X-ray 
and CIV emission from the NPS feature), similarly high levels of emission can
also be found in 
sky-pixels L and O which are thought to be associated with the diffuse gas of the halo.
We note that SiII* is the most abundant ionization state of silicon in the warm and neutral
(weakly ionized) regions
of the ISM. This is in contrast to regions of high CIV emission intensity which generally trace
either collisionally ionized or highly photo-ionized interstellar regions.  There
is some similarity between both emission maps
for positions near to the edge of the NPS (sky-pixels A to F), whereas pixels L and O are
clearly anomalous in their observed high level of SiII* emission. 
Finally, we
note that the level of SiII* from the general halo gas is $\sim$ 2500LU, which is in agreement
with a level of 2430LU recorded by Korpela et al. (\cite{korp06}) towards the NEP region.

\subsection{The AlII emission map}
Figure 5(d) shows the spatial distribution of the AlII (1671\AA) emission. Following the work of
Korpela et al. (\cite{korp06}), we see no evidence for
any measurable emission from the nearby
OIII (1665\AA) emission line in our spectral data. The prior detection
of the OIII line by the lower spectral resolution ($\sim$ 20\AA)
observations of Martin and Bowyer (\cite{martin90}) was probably due to a misidentification
of the detection of emission from the adjacent AlII line.
The AlII ion exists over the ionization potential
range of  6.0 to 18.8eV and under collisional ionization conditions
has its maximum emission at a temperature of $\sim$ 20,000K.
Its relative strength
throughout the presently mapped region suggests that warm-(partially)-neutral
gas is quite pervasive in the halo, thus supporting the notion
that the halo is not isothermal in nature. Although Figure 5(d) shows appreciable levels of AlII 
line emission over
the whole mapped area, the very highest intensities occur in sky-pixels A, C, F and
M. These sky-positions correlate well with those of the denser (and
presumably colder) 21cm HI clouds shown in Figure 4(d).

Finally we note that
Korpela et al. ({\cite{korp06}) observed an AlII emission intensity level of 5610LU towards
the NEP region, which is $\sim$20$\%$ higher than the average level
detected across the whole of the presently mapped region.

\section{FUV Line Ratio Analysis}
\subsection{The High Ions}
In Table 1 we list two important emission line intensity ratios for
the sky-pixels sampled with $\it SPEAR$; that of I(OVI)/I(CIV) and
I(CIV)/I(SiIV). Such ratios
are indicative of the presence of ionized gas with
a temperature $>$ 60,000K. Although much work has been presented on the equivalent ratios from the
corresponding interstellar $\it absorption$ lines of OVI, CIV and SiIV (Zsargo et al. \cite{zsargo03},
Sembach et al. \cite{sembach03}, Indebetouw $\&$ Shull \cite{indeb04b}), very little work
currently exists on measurement of the line ratios recorded in $\it emission$
from the general diffuse IS gas. However,
several theoretical models currently exist that do make
predictions for the expected high ion emission measures based on various assumptions generally
concerning the relative importance of either collisional or photo-ionzation processes
(CLOUDY: Ferland \cite{ferl03}, CHIANTI: Young et al. \cite{young03}). For a comprehensive
list of the major production mechanisms for these high ions we refer the reader to the work of
Indebetouw and Shull (\cite{indeb04a}). At
high galactic latitudes, the most presently favored production mechanism for halo
high ions is that of the galactic fountain (Shapiro $\&$ Field \cite{shap76}), in
which hot gas is liberated into the overlying halo
from expanding supernova remnants and/or stellar bubbles that originate in the galactic
disk. However, the presence of the NPS feature (which is thought to be delineated
by a highly ionized
shocked superbubble or SNR shell), should perhaps argue for a different production mechanism
for the high ions located at galactic longitudes $>$ 285$^{\circ}$. Bearing this in mind,
we might expect the high ion ratios to be different in gas
associated with the (shocked) North Polar Spur region (sky-pixels A - F)
compared with that of the general halo, thus providing us with a diagnostic tool
that can be used to infer differences in the various emission production
mechanisms.

Firstly we note that the I(OVI)/I(CIV) ratio (when measured) gives
values in the relatively restricted range of 1.2 to 2.7, with an
typical error of $\sim$ $\pm$1.0. For sky-positions in which the OVI line
was not detected with confidence (but was detected in CIV), an average value of
I(OVI)/I(CIV) $<$ 1.4 can be assigned to these regions, which is similar to
that of I(OVI)/I(CIV) = 0.98 measured 
for diffuse halo gas near the North Ecliptic Pole
by Korpela et al. (\cite{korp06}).  
Values of the I(OVI)/I(CIV) ratio
in the 0.9 to 3.9 range have been reported for emission from highly ionized
 gas associated with fast moving shocks in SNRs (Raymond et al. \cite{raymond97}, 
 Danforth , Blair $\&$ Raymond \cite{danf01}, Sankrit, Blair $\&$ Raymond \cite{sank03}).
 Thus, perhaps surprisingly, the range of values of the I(OVI)/I(CIV) ratio presently measured
 by us towards the entire region is very similar to that determined for high velocity gas
 associated with SNR shocks.
  
Our range of measured I(OVI)/I(CIV) ratio values
is consistent with that predicted by the
highly ionized halo gas model forwarded by
Shull $\&$ Slavin (\cite{shull94}), which incorporates contributions from
both turbulent mixing layers (Slavin, Shull and Begelman \cite{slavin93})
and isobarically cooling supernova remnants.
Based on an assumed
absorption ratio of N(OVI)/N(CIV) = 3 for a
typical halo sight-line, they predict an emission ratio of I(OVI)/I(CIV) = 1.2 $\pm$0.2
for emitting halo gas with a temperature of log T = 5.3 $\pm$0.3 K, an electron density
of 0.012cm$^{-3}$ and a gas pressure of P/k = 2400cm$^{-3}$.
The actual 
values of high ion absorption measured towards the nearest halo star to our
presently mapped region of the sky
(HD 100340: $\it l$ = 259$^{\circ}$,
$\it b$ = +61$^{\circ}$), is N(OVI)/N(CIV) = 2.8 (Sembach, Savage $\&$ Tripp \cite{semb97}, Zsargo
et al. \cite{zsargo03}), which is very similar to that assumed in the Shull $\&$ Slavin
calculations. We note that our presently measured values of the I(OVI)/I(CIV) ratio 
are inconsistent with the galactic fountain model predictions of
Edgar $\&$ Chevalier (\cite{edgar86}) and Houck $\&$ Bregman (\cite{houck90}) which
respectively predict
ratios an order of magnitude higher and lower than those presently measured.
However,
we shall not pursue specific model predictions in further detail since many of the input
parameters
of  other models (such as
that of Shelton \cite{shelton98}), can be suitably altered to produce results
close to our presently measured values.
We therefore leave a more detailed interpretation of our present
$\it SPEAR$ observations to more qualified theorists.

In order to make a (very) approximate estimate of the the thermal pressure (P/k) of
hot (T = 300,000K) collisionally ionized gas in the halo (of which there are only
a few other measurements),
we use the FUV absorption measurements of Savage et al. (\cite{savage03}) who
observed log N(OVI) = 14.31cm$^{-2}$ towards NGC 4649 (in sky-pixel `C') and
log N(OVI) = 14.73cm$^{-2}$ towards 3C273 (in sky-pixel `J'). If we make the
assumption that
the results from the $\it SPEAR$ OVI emission data (averaged over
64 sq.deg. on the sky) can be meaningfully compared
with the absorption data (gained from pencil beam measurements with $\it FUSE$),
then we may derive
electron density values, n$_{e}$, from
the respective I(OVI)/N(OVI) ratios scaled to those
given
in Shelton et al. (\cite{shelton01}).
From these values of n$_{e}$ we derive estimates
of P/k = 2.1 x 10$^{4}$ cm$^{-3}$ for  hot halo gas in sky-pixel `C'
and P/k =3.3 x 10$^{3}$ cm$^{-3}$ in sky-pixel `J'. We thus find a difference in
hot gas pressure of $\sim$ a factor 7 between the two regions sampled. Recent 
observations of OVI in the galactic halo by Shelton, Sallmen and
Jenkins (\cite{shelton07}) give gas pressures closer to those calculated
for sky-pixel `J'.

Finally, we note that the relatively
restricted range of the I(OVI)/I(CIV) ratio observed by both our $\it SPEAR$ observations
and observations of SNR shocked gas would seem
to argue in favor of a common production mechanism for both ions. Whenever a high level of OVI emission is observed it is always accompanied
by (higher than average) CIV emission with an intensity $>$4300LU. This is probably best explained 
under the assumption that OVI halo emission is somewhat clumpy in nature, consistent with its
 production at interfaces between warm (T = 10$^{3}$ - 10$^{4}$K) and hotter
 (T = 10$^{6}$K) soft X-ray emitting gas. The associated CIV emission at such
 interfaces is thought to occur in the intermediate temperature  (T= 10$^{5}$K) gas, which
seems always present whenever OVI is strongly detected. An alternate explanation
for our results is
that CIV emission may be ubiquitous throughout the halo
with an associated line intensity (of $\sim$ 4000LU) that is fairly constant, and our
present observations merely reflect a
superposition of OVI emission that originates
at the interfaces of random clumps of emitting gas in the halo.

The values of the I(CIV)/I(SiIV) ratio listed in Table 1 span the range 1.6 to 3.2, with a typical measurement error of $\sim$ $\pm$0.8. Korpela et al. (\cite{korp06}) report
a value of I(CIV)/I(SiIV) = 4.1 for the North Ecliptic Pole, which is  an
interstellar region thought to sample
diffuse, warm and ionized halo gas
Although much work
exists on the measurement of the N(CIV)/N(SiIV) ratio derived from halo gas
$\it absorption$ studies (Sembach $\&$ Savage \cite{semb92}), 
the $\it emission$ ratio of I(CIV)/I(SiIV) (like
that of the I(OVI)/I(CIV) ratio)has only been observed
for gas associated with strong, fast moving shocks associated with SNRs or expanding
superbubbles (Raymond et al. \cite{raymond97}, Danforth , Blair $\&$ Raymond \cite{danf01},
Sankrit, Blair $\&$ Raymond \cite{sank03}). In such cases, values of 
the I(CIV)/I(SiIV) ratio in the range 1.9 to 3.9 are typically observed. Thus, again our present observations of a high ion line emission ratio
do not seem to provide us with a sensitive diagnostic that can differentiate
between emission from shocked or the general diffuse and ionized halo gas.  

Since the SiIV ion
is normally associated with lower temperature interstellar regions, as opposed to
the higher temperature and higher ionization lines of NV, CIV and OVI, its contribution
to the total emission from $>$ 10$^{5}$K gas has often been omitted from models that
make line intensity predictions, apart from those regions associated with SNR shocks.
Further work on theoretical predictions are thus clearly required for the more general
case of diffuse and ionized halo gas.

\subsection{The Lower Ions}
Both the SiII* and AlII ions trace mainly diffuse
and warm-(partially)neutral interstellar regions with a gas temperature
$<$ 20,000K. This type of gas seems quite
pervasive in the halo and the distribution of AlII seems well-correlated with that of
21cm HI emitting gas. 
We derive ratio values of I(SiII*)/I(AlII) that span the fairly restricted
range of 0.25 - 0.7 over the whole region,
which is consistent with the value of 0.43 $\pm$0.8 derived by Korpela et al. (\cite{korp06}) for the
North Ecliptic Pole region. In contrast, Kregenow et al. (\cite{kreg06}) derive a significantly higher ratio
of 1.1 for gas associated
with an evolved (hot) thermal interface in the Eridanus Loop superbubble.

We also note from Table 1 that the pixels
with the highest values of CIV emission (A - G) possess values of
I(CIV)/I(AlII) in the range 1.3 - 2.1, whereas sky-pixels with the lowest emission levels
of CIV (K - P) possess ratio values only in the 0.8 - 1.4 range. This could be somewhat significant,
since the former set of pixels can be firmly associated with the hot and highly
ionized shell of the NPS feature,
whereas the latter pixels are thought to sample the general diffuse halo gas that is spatially
distant from the local influence of shocks and/or strong sources of
photo-ionization. 

\section{Conclusion}
We have presented far ultraviolet (FUV:1350 - 1750\AA) spectral imaging
observations recorded with the $\it SPEAR$ satellite
of the interstellar OVI (1032\AA), CIV (1550\AA), SiIV (1394\AA), SiII* (1533\AA) and AlII (1671\AA)
emission lines originating
in a 60$^{\circ}$ x 30$^{\circ}$ rectangular region lying close to the North Galactic Pole.
These data represent the first large area, moderate spatial resolution maps of the
distribution of UV spectral-line emission originating the
both the highly ionized medium (HIM) and the warm ionized medium (WIM)
recorded at high galactic latitudes. Our maps of the OVI, CIV, SiIV and SiII* line emission show the highest intensity levels
being spatially coincident with similarly high levels of soft X-ray emission originating
in the edge of the Northern Polar Spur interstellar feature. However, the distribution of the low
ionization AlII emission does not show a similar spatial correlation, which suggests that
warm-neutral and/or partially ionized gas with a temperature $<$ 20,000K may be quite pervasive at
high galactic latitudes.

The observed emission line intensity ratios
for both high and low ions are discussed, and the variation in these
ratios as a function of
position on the sky is contrasted with the predictions from
current theories concerning the physical state
of the galactic halo. Our data indicates that these various line ratios alone do not provide
us with a clear diagnostic tool to distinguish between the various physical production
mechanisms for both high and low ion states. However,  a comparison between
the spatial distribution and morphology of UV, radio and soft X-ray emission from this
region
does allow us to draw meaningful conclusions concerning the possible production
mechanisms for highly ionized gas at high galactic latitudes.  Our present data favors the hybrid model
of Shull and Slavin (\cite{shull94}),
which incorporates contributions from
both turbulent mixing layers
and isobarically cooling supernova remnants.

Our observed spatial distribution of highly ionized gas near the
North Galactic Pole region
can be best explained under
the assumption that OVI halo emission is somewhat clumpy in nature, consistent with its
 production at interfaces between warm (T = 10$^{3}$ - 10$^{4}$K) and hotter
 (T = 10$^{6}$K) soft X-ray emitting gas. The associated CIV emission at such
interfaces occurs in the intermediate temperature  (T= 10$^{5}$K) gas, which
seems always present (at a higher than average level) whenever OVI is strongly detected.
An alternate explanation
for these observations is
that CIV emission is ubiquitous throughout the halo with
an associated fairly constant level of emission line intensity (of $\sim$ 4000LU),
and therefore our
observations may generally reflect the superposition of spatially
separate OVI emission which originates
at the cloud interfaces of random clumps of high latitude gas.

Finally we note that although these observations represent a
unique data set, our present conclusions
are highly dependent on the limited sensitivity and moderate spectral
resolution of the $\it SPEAR$ instrument. It is clear that more sensitive and
higher spectral resolution emission observations of the FUV sky are required to
answer many of the outstanding problems associated with the presence and physical state
of highly ionized emitting gas in our Galaxy.

\begin{acknowledgements}
$\it SPEAR$/$\it FIMS$ is joint
space project of KASI $\&$ KAIST (Korea) and
U.C. Berkeley (USA), funded by the Korea MOST and
NASA grant NAG5-5355.
We acknowledge the dedicated
team of engineers, technicians, and administrative staff from SSL,
SaTReC and KASI who made this mission possible.
This publication makes use of data products from the SIMBAD database,
operated at CDS, Strasbourg, France. 
\end{acknowledgements}

\end{document}